\documentclass[twocolumn,showpacs,showkeys,preprintnumbers,amsmath,amssymb]{revtex4-1}
\usepackage{bbm}
\usepackage{amsfonts}
\usepackage{graphicx}
\usepackage{amssymb}   
\usepackage{dcolumn}
\usepackage{bm}
\usepackage[colorlinks,citecolor=blue, linkcolor=blue, urlcolor=blue, hyperindex]{hyperref}
\begin{document}
\vspace{2.5cm}

\title{Dual Role of Squeezed-Reservoir in Quantum Phase Synchronization: Boosting and Blockade}
\author{Xing Xiao$^{1}$}
\author{Tian-Xiang Lu$^{1}$}
\author{Wo-Jun Zhong$^{1}$}
\author{Yan-Ling Li$^{2,3}$}
\altaffiliation{liyanling0423@gmail.com}

\affiliation{$^{1}$School of Physics and Electronic Information, Gannan Normal University, Ganzhou, Jiangxi 341000, China\\
$^{2}$School of Information Engineering, Jiangxi University of Science and Technology, Ganzhou, Jiangxi 341000, China\\
$^{3}$Jiangxi Provincial Key Laboratory of  Multidimensional Intelligent Perception and Control, Jiangxi University of Science and Technology, Ganzhou, Jiangxi 341000, China}

\begin{abstract}
 This study explores the dual role of a squeezed reservoir in controlling the quantum phase synchronization of a driven two-level system. We first demonstrate, through a Liouvillian eigen-spectrum analysis, that the squeezed reservoir can induce a stable limit cycle, transforming the passive TLS into a genuine self-sustained oscillator. This enables a qualitative transition from a weak ``forced response" to a robust, high-quality synchronization (or entrainment). This enhancement is characterized not only by a greater degree of phase locking but also by an increased frequency selectivity, manifested as a narrower Arnold tongue. More strikingly, we reveal that the squeezing angle acts as a control parameter to actively suppress synchronization. By tuning this angle, the reservoir can drive the system into a classical mixed state, inducing a quantum synchronization blockade via the quenching of steady-state coherence. Our findings establish squeezed-reservoir engineering as a versatile strategy for actively modulating quantum synchronization, with feasible implementations in circuit quantum electrodynamics.
\end{abstract}
\maketitle

\section{Introduction} 
Synchronization is a fundamental phenomenon observed in various physical, biological, and technological systems where two or more oscillating entities adjust their rhythms through mutual interaction or external driving forces \cite{pikovsky_2001_synchronization}. Synchronization resulting from interactions between entities is typically referred to as spontaneous synchronization \cite{giorgi_2019_transient}, while that resulting from external drives is referred to as forced synchronization (or entrainment) \cite{zhirov_2006_quantum}. Quantum synchronization extends this concept to the realm of microscopic world, where the behavior of systems is governed by the principles of quantum mechanics. In contrast to classical synchronization, quantum synchronization presents a distinctive set of opportunities and challenges.

The opportunities are driven by the potential applications of quantum synchronization in quantum metrology \cite{vaidya_2024_quantum,giorgi_2016_probing}, highly precise timekeeping devices \cite{deutsch_2010_spin}, communication in complex networks \cite{arenas_2008_synchronization,li_2017_quantum}, and encryption communication \cite{spitz_2021_private}. The study of quantum synchronization has also led to many fundamental challenges, including its existence \cite{roulet_2018_synchronizing,parralpez_2020_synchronization}, witness \cite{nadolny_2023_macroscopic,nigg_2018_observing,mahlow_2024_predicting} and measures \cite{giorgi_2012_quantum,mari_2013_measures,ameri_2015_mutual,shen_2023_fisher} of quantum synchronization. For example, a very natural question is, from the perspective of the dimension of Hilbert space, what is the smallest quantum system capable of achieving quantum synchronization? In Ref. \cite{roulet_2018_synchronizing}, the authors claimed that, in order to match the standard paradigm of synchronization, only a qutrit (three-level system) can achieve quantum synchronization, while a two-level system (TLS) cannot due to the absence of a limit cycle. This assertion appears to contradict the known ability of the quantum van der Pol oscillator to achieve synchronization in the deep quantum regime, where only the ground and first-excited states are involved \cite{walter_2014_quantum}. Parra-L\'opez and Bergli \cite{parralpez_2020_synchronization} addressed this apparent discrepancy theoretically by demonstrating that the limit cycle of a single qubit can be established through careful selection of appropriate pure states for constructing mixed states. A trapped-ion experiment tests and verifies this inference by Zhang \emph{et al} \cite{zhang_2023_quantum}. Consequently, it can be concluded that quantum synchronization is indeed achievable for a single qubit. Furthermore, in this work, we explicitly bridge this minimal two-level paradigm with standard continuous-variable synchronization by deriving it from the deep quantum limit of a quantum Stuart-Landau oscillator (QSLO). As we know, quantum noise represents a significant challenge in the pursuit of quantum synchronization. Some researches show that noise can also induce mutual synchronization in many-body system \cite{schmolke_2022_noiseinduced,tao_2024_noiseinduced,song_2024_spinnoiseinduced}. However, in the deep quantum region, even minor fluctuations in quantum states can result in the system failing to achieve stable synchronization with external drive signals \cite{giorgi_2013,bellomo_2017}. While quantum synchronization is often studied in open systems subject to dissipation and an external drive \cite{roulet_2018_synchronizing,parralpez_2020_synchronization,zhang_2023_quantum,schmolke_2022_noiseinduced,tao_2024_noiseinduced,song_2024_spinnoiseinduced,giorgi_2013,bellomo_2017}, it is noteworthy that synchronization phenomena can also emerge intrinsically from the conservative, nonlinear dynamics of isolated many-body systems. For instance, following a quantum quench in the Bose-Hubbard model, it has been shown that the system can dynamically relax to a state of ``$\pi$-synchronization," where oscillators form clusters with a relative phase of $\pi$ \cite{pizzi_2019}. This highlights the rich variety of mechanisms that can lead to synchronization in the quantum realm.

As we all know that the coupling of a quantum system to its environment is typically a source of decoherence~\cite{breuer_2006_the}, the paradigm of reservoir engineering offers a powerful alternative: structuring the environment to control the system's dynamics~\cite{zeytinolu_2017_engineering,zippilli_2021_dissipative,bai_2021_generating}. Previous studies have indeed explored the interplay between squeezing and synchronization~\cite{manzano_2013_synchronization,weiss_2017_quantumcoherent,shen_2023_enhancing,sonar_2018_squeezing}. However, these approaches primarily focus on squeezing the system's internal degrees of freedom, such as its initially squeezed state~\cite{manzano_2013_synchronization,weiss_2017_quantumcoherent,shen_2023_enhancing} or a direct drive~\cite{sonar_2018_squeezing}, leaving the role of a continuously coupled, non-trivially engineered environment largely unexplored. This motivates our central question: Can a squeezed reservoir be used to actively control and manipulate quantum phase synchronization?

In this work, we answer this question and demonstrate that a squeezed reservoir acts as a dual-function tool. On one hand, it can moderately enhance the quality and selectivity of quantum phase synchronization, leading to highly localized Arnold tongues, which indicates a stronger frequency selectivity and phase-locking resilience than those in conventional environments. On the other hand, and more strikingly, we show that by tuning the squeezing angle $\Phi$, a previously overlooked control parameter, the reservoir can induce a near-complete quantum synchronization blockade~\cite{mok_2020_synchronization,lrch_2017_quantum,solanki_2023_symmetries,wang_2023_absence}, actively suppressing the system's phase-locking. We further propose a feasible experimental implementation in a circuit quantum electrodynamics (QED) architecture. Our findings establish squeezed-reservoir engineering as a potent strategy for both boosting and quenching quantum phase synchronization, offering a new degree of control for quantum technologies.

The structure of the paper is organized as follows. In Sec. \ref{sec2}, we provide a concise mathematical framework characterizing a TLS embedded within a squeezed reservoir, establish its origin from a strongly non-linear bosonic oscillator, and derive its stationary solution. In Sec. \ref{sec3} we demonstrate the presence of stable limit cycles in the TLS within the squeezed-vacuum reservoir through both transient phase-space trajectories and rigorous algebraic Liouvillian analysis, which are absent in the vacuum reservoir. Subsequently, in Sec. \ref{sec4}, we discuss the primary impact of the squeezed reservoir on quantum phase synchronization, namely boosting, as well as show the distinctive properties of Arnold tongues. Sec. \ref{sec5} focuses on revealing the secondary influence of the squeezed reservoir on quantum phase synchronization, i.e, synchronization blockade, by varying the squeezing angle. Finally, we present an experimental discussion in Sec. \ref{sec6} and summarize our findings in Sec. \ref{sec7}.

\begin{figure*}
  \includegraphics[width=0.99\textwidth]{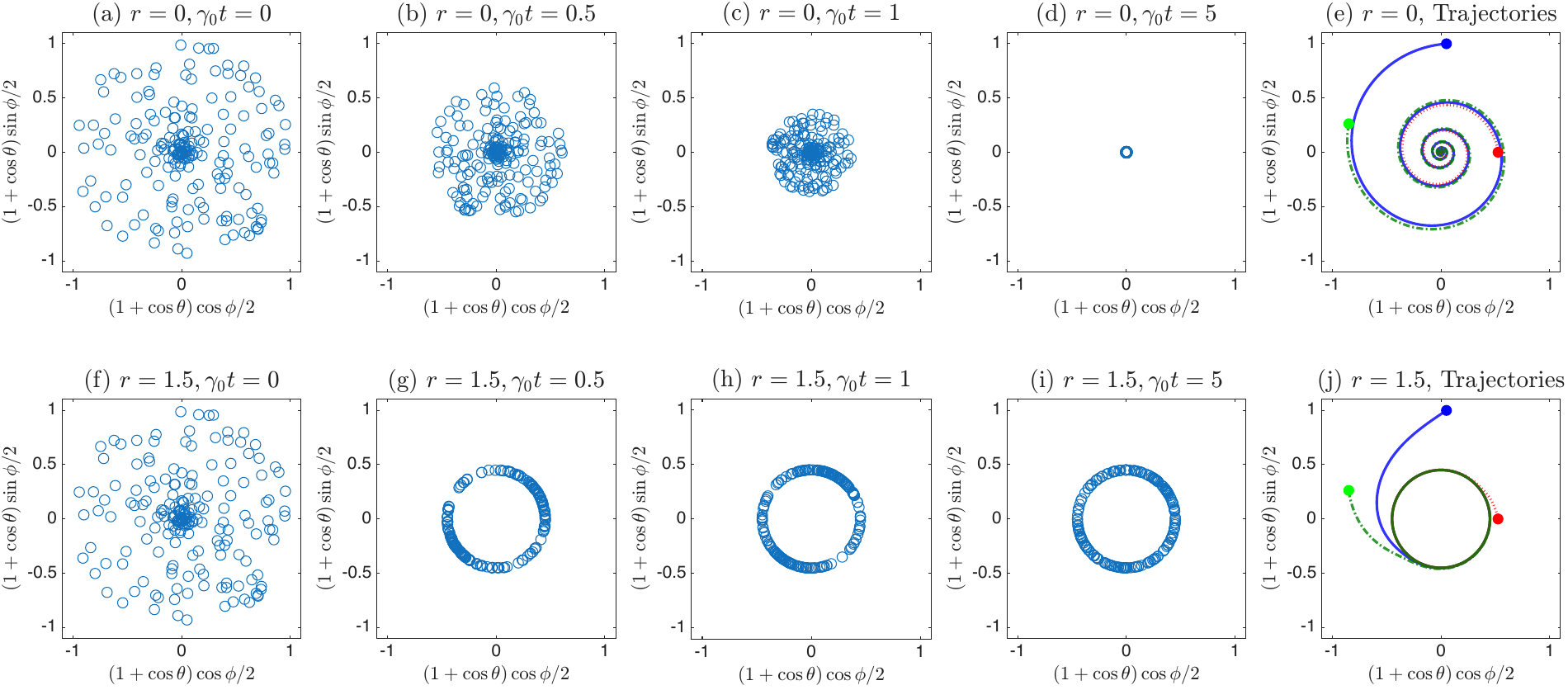}
\caption{(color online) Simulations of classical trajectories in phase space for different reservoirs are presented: panels (a)-(e) correspond to a vacuum reservoir, while panels (f)-(j) correspond to a squeezed-vacuum reservoir. In panels (a)-(e) and (f)-(i), each circle denotes the evolution of one random initial state. From left to right, the positions of 200 random initial states in phase space at various times ($\gamma_{0}t=0,\ 0.5,\ 1,\ 5$) are represented. Panels (e) and (j) show the transient time evolution of the un-driven system prior to reaching the steady state for three random initial states in vacuum and squeezed reservoirs, respectively. The cycle depicted in panels (h) and (j) illustrates a limit cycle.} The remaining parameters are set to $n=0$, $\Phi=0$, $\Delta=0$ and $\epsilon=0$.
\label{fig1}       
\end{figure*}

\section{Model} 
\label{sec2}
\label{sec2}
 {We consider a TLS with transition frequency $\omega_{0}$ coupled to a squeezed-thermal reservoir which is mimicked by a lossy waveguide driven by a broadband squeezed laser~\cite{kono_2017}. 
The total Hamiltonian of the system and reservoir is given by $H_{\rm tot} = H_S + H_R + H_I$ ($\hbar=1$):
\begin{align}
\label{eq1}
    \hat{H}_{\rm tot}=\frac{\omega_0}{2}\hat{\sigma}_{z}+\sum_k\omega_k \hat{a}_k^\dagger \hat{a}_k+ \sum_k (g_k \hat{a}_k \hat{\sigma}_{+}+ \text{H.c.}),
   \end{align}
with $\hat{a}_k^\dagger$ and $\hat{a}_k$ being the creation and annihilation operators for the $k$-th mode of the reservoir with frequency $\omega_k$. $\hat{\sigma}_{z}=|e\rangle\langle e|-|g\rangle\langle g|$ is the Pauli operators. The coupling strength between the TLS and the $k$-th mode is $g_{k}$. For a squeezed-thermal reservoir, the density operator of the reservoir is given by:
\begin{equation}
\label{eq2}
    \rho_R = \hat{S}(\xi) \rho_{\text{th}}(T) \hat{S}^\dagger(\xi),
\end{equation}
 {where $\hat{S}(\xi) = \exp\left[\frac{1}{2} \sum_k (\xi^* \hat{a}_k^2 - \xi (\hat{a}_k^\dagger)^2)\right]$ is the squeezing operator, $\rho_{\text{th}}(T)$ is the thermal density operator at temperature $T$, and $\xi = r e^{i\Phi}$ is the complex squeezing parameter with $r$ the squeezing strength and $\Phi$ the squeezing angle.}}

 {The dynamics depend on the two-time correlation functions of the reservoir. For a squeezed-thermal state, these correlations are non-zero only for specific frequency relations. The relevant quantities are~\cite{gardiner_1986}:
\begin{eqnarray}
\label{eq3}
\langle\mathcal{A}^{\dagger}(t)\mathcal{A}(t')\rangle&=&\gamma_{0}N\delta(t-t'),\nonumber\\
\langle\mathcal{A}(t)\mathcal{A}^{\dagger}(t')\rangle&=&\gamma_{0}(N+1)\delta(t-t'),\\
\langle\mathcal{A}(t)\mathcal{A}(t')\rangle&=&\gamma_{0}M^{*}e^{-2i\omega_{0}t}\delta(t-t'),\nonumber\\
\langle\mathcal{A}^{\dagger}(t)\mathcal{A}^{\dagger}(t')\rangle&=&\gamma_{0}M e^{2i\omega_{0}t}\delta(t-t'),\nonumber
\end{eqnarray}
where $\mathcal{A}(t)=\sum_{k}g_{k}\hat{a}_{k}\exp({-i\omega_{k}t})$ is the reservoir operator in the interaction picture. $\gamma_{0}$ denotes the spontaneous emission rate of the TLS. The effective mean number of photons observed by the system is expressed as $N = n \cosh(2r) + \sinh^{2}(r)$, where $n = (e^{\hbar \omega_k / k_B T} - 1)^{-1}$ represents the mean number of thermal photons in the reservoir with Boltzmann constant $k_B$. Additionally, the parameter $M=-\frac{1}{2}\sinh2r e^{i\Phi}(2n+1)$ characterizes the strength of the correlations induced by squeezing.}

In order to synchronize the TLS with a weak signal, we use a semiclassical drive with frequency $\omega_{L}$ and strength $\epsilon$. In the rotating-wave approximation, the signal Hamiltonian will be written as 
\begin{equation}
\label{eq4}
\hat{H}_{\rm signal}=i\frac{\epsilon}{4}(e^{i\omega_{L} t}\hat{\sigma}_{-}-e^{-i\omega_{L} t}\hat{\sigma}_{+}).
\end{equation}
Moving to a frame rotating with the drive frequency $\omega_{L}$  {by transformation $\hat{T}_{\omega_{L}}=\exp(i\omega_{L}\hat{\sigma}_{z}t/2)$} and tracing out the reservoir, the final master equation for the TLS takes the Lindblad form \cite{bellac_2012_quantum}
\begin{eqnarray}
\label{eq5}
\frac{d\rho}{dt}&=&-\frac{i}{2}[\Delta\hat{\sigma}_{z}+\epsilon\hat{\sigma}_{y},\rho]-\gamma_{0}M\hat{\sigma}_{+}\rho\hat{\sigma}_{+}-\gamma_{0}M^{*}\hat{\sigma}_{-}\rho\hat{\sigma}_{-}\nonumber\\
&&-\frac{1}{2}\gamma_{0}(N+1)(\hat{\sigma}_{+}\hat{\sigma}_{-}\rho+\rho\hat{\sigma}_{+}\hat{\sigma}_{-}-2\hat{\sigma}_{-}\rho\hat{\sigma}_{+})\nonumber\\
&&-\frac{1}{2}\gamma_{0}N(\hat{\sigma}_{-}\hat{\sigma}_{+}\rho+\rho\hat{\sigma}_{-}\hat{\sigma}_{+}-2\hat{\sigma}_{+}\rho\hat{\sigma}_{-}),
\end{eqnarray}
where $\Delta=\omega_{L}-\omega_{0}$ is the detuning between the frequency of the drive and the TLS. 

To provide a physical context for the origin of the quantum limit cycle within this minimal two-level manifold and to establish a connection with established synchronization paradigms, it is instructive to derive Equation (\ref{eq5}) from the deep quantum limit of a QSLO. We have demonstrated that our squeezed TLS model can be obtained from a driven QSLO by incorporating a Kerr term in the deep quantum limit \cite{sonar_2018_squeezing,Lim2024}; further details can be found in Appendix \ref{sec:QSLO}.

Here, we use the Bloch representation of a TLS to solve the master equation
\begin{equation}
\label{eq6}
\rho=\frac{1}{2}(1+\boldsymbol{r} \cdot \hat{\boldsymbol{\sigma}}),
\end{equation}
where $\boldsymbol{r}=\left(r_{x}, r_{y}, r_{z}\right)^{\mathrm{T}}$ denotes the Bloch vector and $\hat{\boldsymbol{\sigma}}=\left(\hat{\sigma}_{x}, \hat{\sigma}_{y}, \hat{\sigma}_{z}\right)$. The stationary solutions of Eq.~(\ref{eq5}) can be obtained analytically
\begin{eqnarray}
\label{eq7}
r_{x}=\frac{2\gamma_{0}\epsilon\big[\gamma-2\gamma_{0}\text{Re}(M)\big]}{\gamma\big[4(\gamma_{0}^{2}|M|^{2}-\Delta^{2})-\gamma^{2}\big]+2\epsilon^2\big[2\gamma_{0}\text{Re}(M)-\gamma\big]},\\
\label{eq8}
r_{y}=\frac{4\gamma_{0}\epsilon\big[\Delta+\gamma_{0}\text{Im}(M)\big]}{\gamma\big[4(\gamma_{0}^{2}|M|^{2}-\Delta^{2})-\gamma^{2}\big]+2\epsilon^2\big[2\gamma_{0}\text{Re}(M)-\gamma\big]},\\
\label{eq9}
r_{z}=\frac{-\gamma_{0}[4(\gamma_{0}^{2}|M|^{2}-\Delta^{2})-\gamma^{2}]}{\gamma\big[4(\gamma_{0}^{2}|M|^{2}-\Delta^{2})-\gamma^{2}\big]+2\epsilon^2\big[2\gamma_{0}\text{Re}(M)-\gamma\big]},
\end{eqnarray}
where $\gamma=\gamma_{0}(2N+1)$.

\section{Limit cycle} 
\label{sec3}
Fundamentally, the emergence of synchronization is predicated on the existence of a limit cycle \cite{pikovsky_2001_synchronization,roulet_2018_synchronizing}. For the TLS coupled to a squeezed reservoir, we will show that it has a valid limit cycle in the Bloch space $\{\theta,\phi\}$. In order to track the `` trajectories'' of the TLS, it is necessary to calculate the time evolution of both $\theta$ and $\phi$. These are determined by the following equations
\begin{eqnarray}
\label{eq10}
\frac{d\theta}{dt}&=&\gamma_0\csc\theta+\gamma\cot\theta-2\epsilon\cos(\omega_{L}t+\frac{\pi}{2})\sin\phi,\\
\label{eq11}
\frac{d\phi}{dt}&=&\omega_{0}+\gamma_0\text{Im}(M)\cos2\phi+\gamma_0\text{Re}(M)\sin2\phi\nonumber\\
&&-2\epsilon\cos(\omega_{L}t+\frac{\pi}{2})\cos\phi\cot\theta.
\end{eqnarray}
When the external drive is absent, the phase $\phi$ is fully free, but we can obtain the steady-state solution of $\theta$
\begin{equation}
\label{eq12}
\theta_{s}=\arccos\left(\frac{-1}{2N+1}\right).
\end{equation}
It is noteworthy that when $N=0$ (i.e., $n=0$ and $r=0$), the steady-state of the TLS is identified as $|0\rangle$ when $\theta_s=\pi$. Such a state is unable to provide a valid limit cycle for synchronization, given that it represents a single point.  
Nevertheless, if $r\neq0$, a finite value of $N$ may represent a valid limit cycle.

In order to obtain an intuitive understanding of the limit cycle, we simulate the ``classical trajectories'' of the TLS in the normalized rectangular coordinates $\{\frac{(1+\cos\theta)\cos\phi}{2},\frac{(1+\cos\theta)\sin\phi}{2}\}$. {Figs.~\ref{fig1}(a)-(d) illustrate the transient evolution of the system under a vacuum reservoir (i.e., $n=0$, $r=0$) with 200 randomly generated initial states. It can be demonstrated that all circles eventually converge to the point with coordinates $(0,0)$, which corresponds to the ground state $|0\rangle$. This is a straightforward consequence of the fact that, in the absence of any gain, a dissipative TLS inevitably decays to the ground state. In contrast, when a squeezed-vacuum reservoir is employed (i.e., $n=0$, $r=1.5$), all the states evolve onto a cycle because the $\theta_s$ is finite and the phase $\phi$ is fully free in the absence of external drive, as shown in Figs.~\ref{fig1}(f)-(i). The radius of the limit cycle is $R_s=(1+\cos\theta_s)/2$.
Figs.~\ref{fig1}(e) and (j) show the transient time evolution of the un-driven system prior to reaching the steady state for three random initial states in vacuum and squeezed reservoirs, respectively. As clearly demonstrated, rather than decaying trivially to a single point in the vacuum reservoir, which would be the case for a passive damped oscillator. All trajectories spiral into and continuously wrap around a macroscopic, closed circular attractor in the squeezed reservoir. This non-zero robust basin of attraction confirms that the squeezed reservoir structurally elevates the qubit into a genuine self-sustained limit-cycle oscillator.
 The linear stability analysis of the un-driven ($\epsilon = 0$) system is discussed in Appendix \ref{sec:appendix_stability}. While these mean-field trajectories provide an intuitive semiclassical picture, they inherently neglect quantum fluctuations.

It is important to note that since the master equation (\ref{eq5}) is described in a frame rotating at the drive frequency $\omega_L$, its stationary solution ($d\rho/dt = 0$) corresponds to the long-term state of the system synchronized with the external drive. In this rotating frame, a stable limit cycle in the non-rotating frame (i.e., a state oscillating with a definite amplitude at frequency $\omega_L$ over time) manifests as a stable fixed point. The relation between the limit cycle in non-rotating frame and the time-independent point in rotating frame is characterized by the transformation $\rho '=\hat{T}_{\omega_{L}}\rho\hat{T}_{\omega_{L}}^{\dagger}$. We can obtain the time-dependent Bloch vector in the non-rotating frame: $r'_{x}=r_{x}\cos\omega_{L}t-r_{y}\sin\omega_{L}t$, $r'_{y}=r_{x}\sin\omega_{L}t+r_{y}\cos\omega_{L}t$ and $r'_{z}=r_{z}$. This vector corresponds to spin precession with a stable amplitude, independent of the initial state. In Refs. \cite{parralpez_2020_synchronization, zhang_2023_quantum}, the authors explained that the stationary mixed state can be interpreted as a probabilistic mixture of pure states. While the overall ensemble remains stationary, each individual pure state periodically precesses, thereby providing a valid limit cycle. 

Although the above argument has provided a physical insight of the existence of limit cycle, we can further confirm this from an algebraic perspective. 
The evidence of a quantum limit cycle lies in the dynamical eigen-spectrum of the Liouvillian super-operator. According to the algebraic theory of quantum synchronization  \cite{buca_2022_algebraic}, self-sustained oscillations emerge when the Liouvillian possesses eigen-modes with vanishingly small real parts (corresponding to extremely slow decoherence) and non-zero imaginary parts (corresponding to coherent oscillations). To demonstrate this algebraically, we consider the un-driven system ($\epsilon = 0$) in the interaction picture. The dynamics are entirely governed by the dissipation super-operator. By projecting the density matrix onto the Pauli basis $\hat{\sigma} = (\hat{\sigma}_x, \hat{\sigma}_y, \hat{\sigma}_z)$, we obtain the decay rates of the three orthogonal components of the Bloch vector, which correspond to the absolute values of the non-zero Liouvillian eigenvalues $\lambda_i$. For clarity, we set the squeezing angle $\Phi=0$, yielding a real and negative squeezing parameter $M = -|M|$. The corresponding Liouvillian eigenvalues are analytically given by:
\begin{align}
\lambda_x^{\rm sq} &= -\gamma_0(N + 1/2 - |M|),\nonumber \\
\lambda_y^{\rm sq} &= -\gamma_0(N + 1/2 + |M|),\\
\lambda_z^{\rm sq} &= -\gamma_0(2N + 1).\nonumber
\end{align}

In a standard vacuum reservoir without squeezing ($N=0, M=0$), these eigenvalues reduce to $\lambda_x^{\rm vac} = \lambda_y^{\rm vac} = -\gamma_0/2$ and $\lambda_z^{\rm vac} = -\gamma_0$. All eigen-modes possess macroscopic negative real parts on the order of the spontaneous emission rate $\gamma_0$, meaning the system acts as a trivial point attractor. Any initial quantum state rapidly and isotropically collapses to the ground state, fundamentally precluding the formation of a limit cycle.
However, the dynamic landscape undergoes a topological transition when a highly squeezed reservoir is introduced. In the deep quantum regime with strong squeezing ($r \gg 1$, hence $N \gg 1$), we can asymptotically expand the correlation strength $|M| = \sqrt{N(N+1)} \approx N + 1/2 - 1/(8N)$. Substituting this expansion into the eigen-spectrum yields a separation of timescales:
\begin{align}
\lambda_x^{\rm sq} &\approx -\frac{\gamma_0}{8N},\nonumber \\
\lambda_y^{\rm sq} &\approx -2\gamma_0 N, \\
\lambda_z^{\rm sq} &\approx -2\gamma_0 N.\nonumber 
\end{align}

This mathematical structure provides an evidence for a quantum limit cycle. The squeezed reservoir induces severe quantum fluctuations along the $y$ and $z$ axes, forcing the system to undergo rapid amplitude relaxation ($|\lambda_y^{\rm sq}|, |\lambda_z^{\rm sq}| \propto N$). In stark contrast, the phase diffusion rate along the $x$-axis is significantly suppressed ($|\lambda_x^{\rm sq}| \propto 1/N$), asymptotically approaching zero. This extreme dissipation anisotropy carves out a highly slow manifold in the state space.
When transformed back to the Schrodinger picture, this extremely long-lived eigen-mode naturally acquires an imaginary part $\pm i\omega_0$. The emergence of an eigen-mode with an infinitesimally small real part and a macroscopic non-zero imaginary part strictly fulfills the algebraic criteria for a metastable quantum limit cycle. It is this protected slow manifold that transforms the passive two-level system into a genuine self-sustained oscillator. This mechanism intrinsically explains why the squeezed reservoir can provide a low-dissipation channel for a weak external drive to achieve robust quantum phase synchronization, facilitating the qualitative transition from a forced response to genuine entrainment, as we will discuss below.

\begin{figure}
  \includegraphics[width=0.5\textwidth]{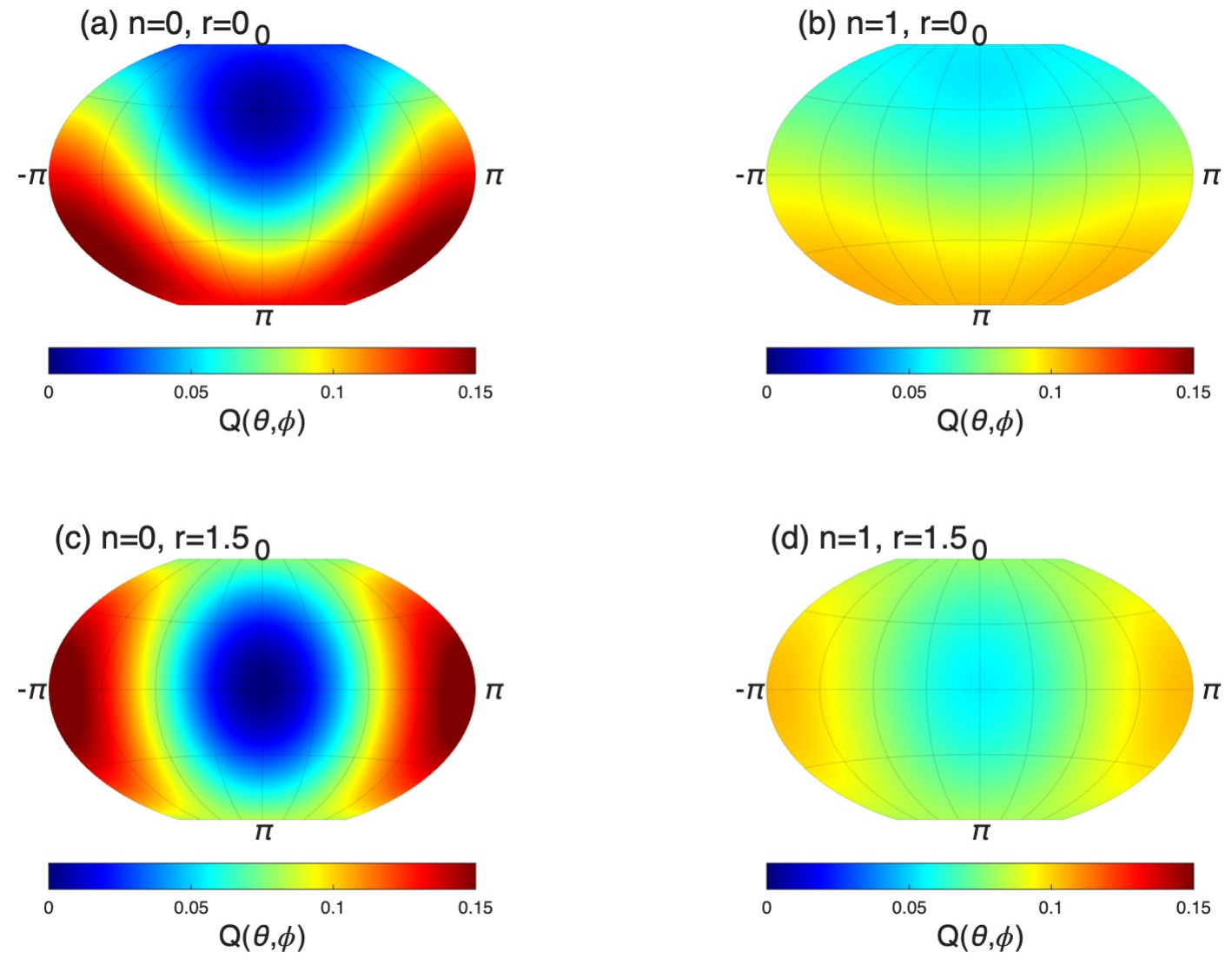}
\caption{(color online) The steady-state $Q$-function of TLS in different reservoirs: (a) vacuum reservoir: $n=0$, $r=0$. (b) thermal reservoir: $n=1$, $r=0$. (c) squeezed-vacuum reservoir: $n=0$, $r=1.5$. (d) squeezed-thermal reservoir: $n=1$, $r=1.5$. The other parameters are $\Phi=0$, $\Delta=0$ and $\epsilon/\gamma_0=0.5$.} 
\label{fig2}       
\end{figure}

\section{Boosting synchronization by squeezed reservoir} 
\label{sec4}

\subsection{Phase distribution: $Q$-function}
We now proceed to examine the phenomenon of quantum phase synchronization in the context of an external signal. As a preliminary exploration of reservoir-engineered quantum phase synchronization, we employ the Husimi $Q$-function as a means of evaluating the phase preference of the TLS in terms of the coherent-spin states. The Husimi $Q$-function is one of the simplest distributions of quasiprobability in phase space. It is defined for any TLS state, represented by the density operator $\rho$, as follows:
\begin{equation}
\label{eq13}
Q(\theta,\phi)=\frac{1}{2\pi}\langle \theta, \phi |\rho|\theta, \phi \rangle
\end{equation}
where $|\theta, \phi \rangle=\cos\frac{\theta}{2}|1\rangle+\sin\frac{\theta}{2}\exp(i\phi)|0\rangle$ are coherent-spin states which are defined by the eigenstates of the operator $\hat{\boldsymbol{\sigma}}\cdot{\boldsymbol{n}}$ with $\boldsymbol{n}=(\sin\theta\cos\phi,\sin\theta\sin\phi,\cos\theta)$. 
According to Eqs. (\ref{eq7})-(\ref{eq9}), the exact formulation of $Q$-function is
\begin{equation}
\label{eq14}
Q(\theta,\phi)=\frac{1+(r_{x}\cos \phi+r_{y}\sin \phi) \sin\theta+r_{z}\cos\theta}{4\pi}.
\end{equation}

We begin by qualitatively analyzing the phase synchronization through the Husimi $Q$-function, which visualizes the phase preference of the TLS in the coherent-spin state space. Figure 2 depicts the Husimi $Q$-function $Q(\theta, \phi)$ for the steady state using the Winkel tripel projection of a sphere. When evaluating these phase space distributions, a fundamental distinction must be drawn. Unlike classical oscillators, where synchronization sharpness can be arbitrarily high, a single qubit's phase localization is strictly limited by fundamental quantum fluctuations. Because the two-level Hilbert space is bounded, the $Q$-function for any spin state intrinsically possesses a macroscopic zero-point broadening. Therefore, the visual peakedness cannot exhibit an infinitely sharp spike even under perfect phase locking. In this deeply quantum regime, the hallmark of synchronization is not a classical delta-peak, but rather the emergence of a well-defined phase preference.

It is crucial to recognize that the $r=0$ and $r>0$ scenarios represent two fundamentally different dynamical regimes. The $r=0$ case depicted in Fig.~\ref{fig2}(a) corresponds to a purely dissipative system which, as established in Sec. \ref{sec3}, lacks a stable limit cycle. The $Q$-function is broadly distributed, indicating a significant degree of phase uncertainty, although it is peaked around the expected phase value of $\phi=\pi$ \cite{xiao_2023_classicaldrivingassisted,pizzi_2019}. Its behavior under the drive is therefore not synchronization in the strict sense, but rather a forced response. The external signal drives the passive system, forcing it to oscillate and thereby creating a phase preference. However, this response is weak and susceptible to noise due to the system's intrinsic tendency to decay, resulting in a broad and diffuse peak in the $Q$-function. This can be confirmed in Fig.~\ref{fig2}(b) where thermal fluctuations are introduced by setting $n=1$.

 In contrast, the $r>0$ case presented in Fig.~\ref{fig2}(c) represents true synchronization (or entrainment). The squeezed reservoir transforms the TLS into a genuine ``self-sustained oscillator'' possessing a stable limit cycle. The external drive then acts to phase-lock this intrinsic oscillation. Because this process involves entraining an active oscillator, the resulting phase-locking is strong and resilient to thermal fluctuations, which is manifested as a sharp and highly localized peak in the $Q$-function. Fig.~\ref{fig2}(d) shows that the beneficial effect of squeezing persists and even dominates in the presence of thermal noise. When the TLS interacts with a squeezed-thermal reservoir, the resulting phase distribution is substantially sharper and more localized than in the thermal-only case.

\subsection{Synchronization measure: $S$-function}

A qualitative understanding of quantum phase synchronization has been obtained from the perspective of the $Q$-function.
The objective of the subsequent analysis is to examine the phase synchronization in quantitative terms. In order to quantify the degree of quantum phase synchronization, we adopt the approach proposed in Refs. \cite{roulet_2018_synchronizing,parralpez_2020_synchronization,xiao_2023_classicaldrivingassisted} whereby the $S$-function is utilized.
The synchronization measure $S(\phi)$ is defined by integrating over the angular variable $\theta$:
\begin{equation}
S(\phi)=\int_{0}^{\pi}d\theta\sin\theta Q(\theta,\phi)-\frac{1}{2\pi},
\label{eq15}
\end{equation}
which vanishes everywhere if and only if there is no phase synchronization between the TLS and the external signal.

In our model, the synchronization measure $S(\phi)$ reduces to
\begin{equation}
S(\phi)=\frac{1}{8}(r_{x}\cos \phi +r_{y}\sin \phi).
\label{eq16}
\end{equation}
Using the synchronization measure $S(\phi)$, we can identify the parameter regions in which the synchronization takes place.

\begin{figure}
  \includegraphics[width=0.5\textwidth]{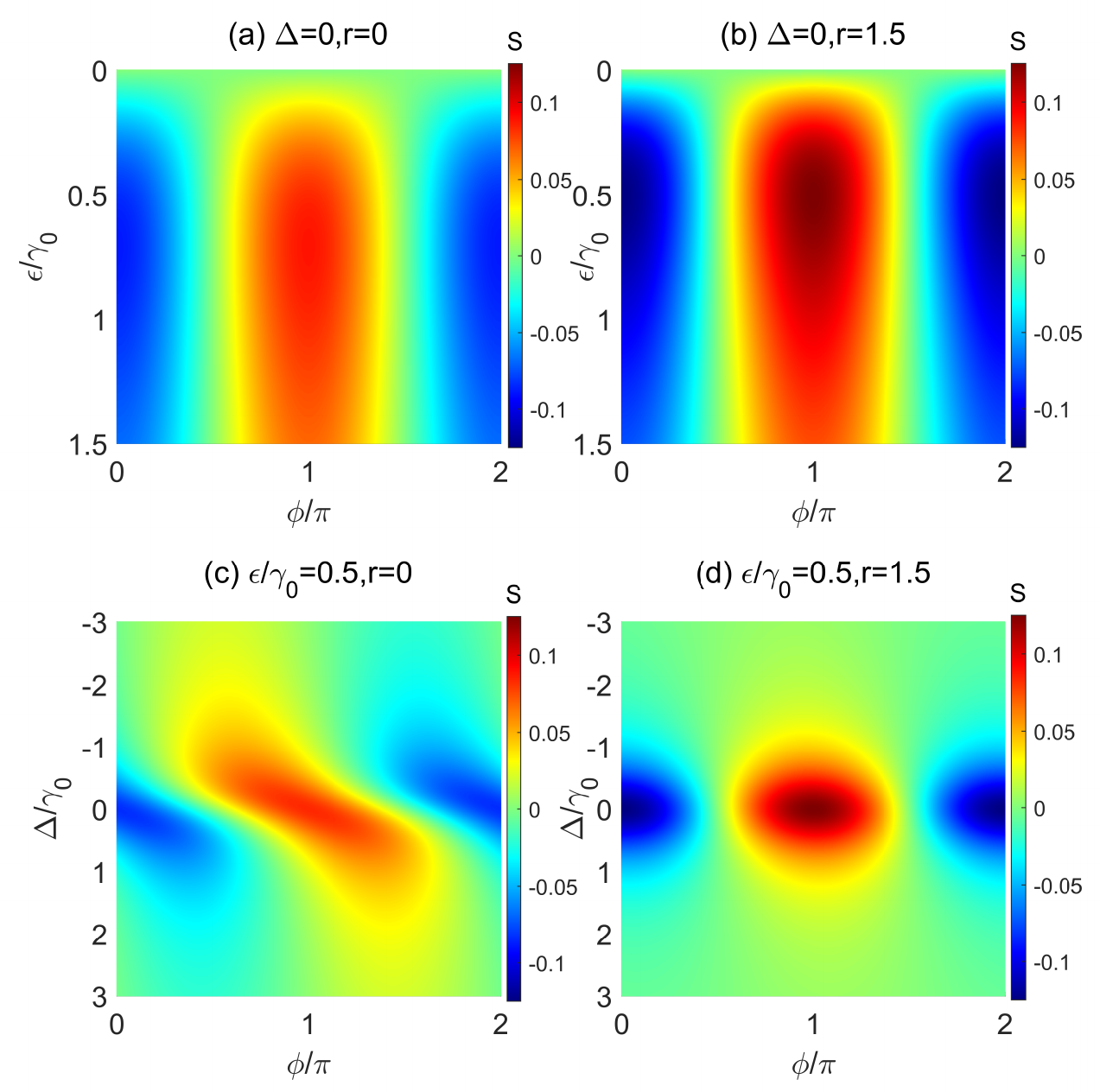}
\caption{(color online) The distribution of synchronization measure $S(\phi)$ in different parameter regions. (a) $S(\phi)$ as a function of the signal strength $\epsilon$ and the phase of TLS in the
vacuum reservoir: $n=0$, $r=0$, and (b) in the squeezed-vacuum reservoir: $n=0$, $r=1.5$. (c) $S(\phi)$ as a function of the detuning $\Delta$ and the phase of TLS in the
vacuum reservoir: $n=0$, $r=0$, and (d) in the squeezed-vacuum reservoir: $n=0$, $r=1.5$. The other parameters are the same as Fig.~\ref{fig2}.}
\label{fig3}       
\end{figure}

The initial factor that we examine is the impact of signal strength $\epsilon$ by fixing the detuning $\Delta=0$. This behavior is illustrated in Figs.~\ref{fig3}(a) and \ref{fig3}(b), which demonstrate how the distribution of $S(\phi)$ is affected by the signal strength, both in the absence and presence of squeezed-reservoir engineering. It was found that a gradual increase in signal strength is beneficial for phase synchronization. This is due to the fact that when the system is subjected to both noise and a drive signal, a stronger signal can reduce the tendency of phase diffusion induced by the fluctuations of the noise. However, a signal of excessive strength would drive the TLS out of the synchronization regime. This can be understood as follows: in accordance with the standard theory of synchronization, the formation of the limit cycle is determined by the subtle relationship between gain and loss of the system. When the drive signal is excessively strong, it compromises the stability of the limit cycle, thereby precluding the occurrence of phase synchronization. Indeed, the signal strength should be moderate that ensures it only influences the phase of the TLS but does not affect the shape of the limit cycle. 

One can find the optimal strength of the signal by solving the equation ${\partial S(\phi)}/{\partial\epsilon}=0$. The specific expression of $\epsilon_{\rm opt}$ is too complex to be written out with a clear physical insight, but under the parameters illustrated in Fig.~\ref{fig3}(a), this expression can be simplified to $\epsilon_{\rm opt}=\gamma/\sqrt{2}$ in the vacuum reservoir. The optimal value of $\epsilon$ yields to $0.707\gamma_0$.
When we take into account the squeezing effect of the reservoir on the synchronization measure, the optimal strength is $\epsilon_{\rm opt}=\sqrt{\gamma^2-2\gamma\gamma_{0}|M|}/\sqrt{2}$ under the parameters illustrated in Fig.~\ref{fig3}(b), which is equal to $0.5\gamma_0$. It is evident that the quantum phase synchronization exhibits a tendency towards greater localization and pronounced manifestation in the squeezed-vacuum reservoir. 

The second factor that we examine is the impact of detuning $\Delta$ by fixing the signal strength $\epsilon/\gamma_0=0.5$. Fig.~\ref{fig3}(c) shows how the preferred phase is dragged away from the reference phase $\phi=\pi$ by the detuning. As the absolute value of the detuning is increased, the synchronization measure $S(\phi)$ is weakened and will be averaged out in the far-off-resonance regime.
It is noteworthy that the introduce of a squeezed reservoir will result in the phase preference of the system becoming a regular normal distribution around the reference phase, which leads to a ``cleaner'' phase synchronization, as shown in Fig.~\ref{fig3}(d). 

\subsection{Synchronization region: Arnold tongue and the algebraic closure}

\begin{figure}
  \includegraphics[width=0.5\textwidth]{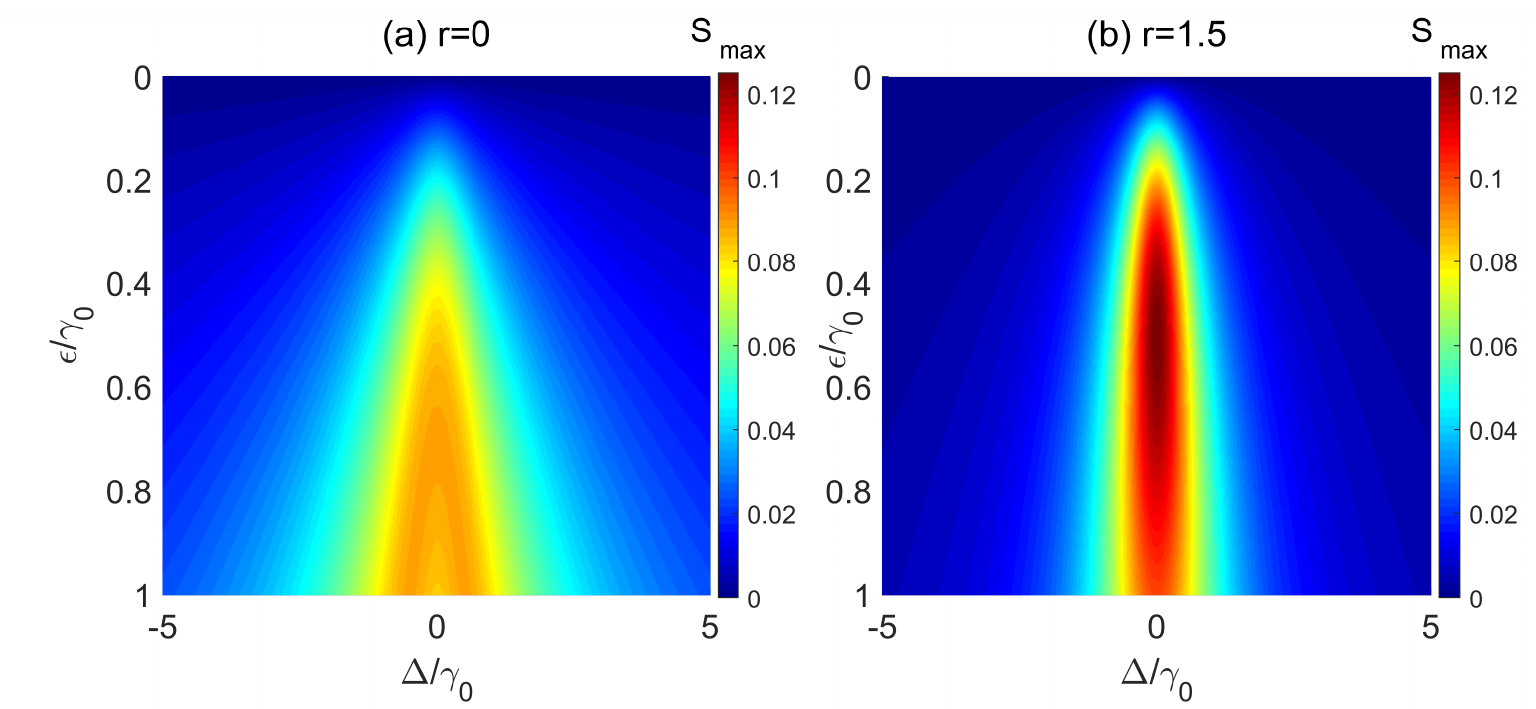}
\caption{(color online) Schematic representation of Arnold tongues. (a) $S_{\rm max}$ as a function of the signal strength $\epsilon$ and the detuning $\Delta$ in the thermal reservoir ($n=1$, $r=0$) and (b) in the squeezed-thermal reservoir ($n=1$, $r=1.5$). The other parameters are the same as Fig.~\ref{fig2}.}
\label{fig4}       
\end{figure}

It is enlightening to plot the behaviors of $S_{\rm max}$ which is the maximum of $S(\phi)$ varying $\phi$ from 0 to $2\pi$. $S_{\rm max}$ captures the synchronization regardless of the phase preference in a specific orientation. Fig.~\ref{fig4}(a) depicts $S_{\rm max}$ as a function of the signal strength $\epsilon$ and the detuning $\Delta$ in the thermal reservoir without squeezing. It determines the region in the $(\epsilon, \Delta)$ parameter space corresponding to the quantum phase-synchronized regime, universally known as the Arnold tongue \cite{reichhardt_1999_phase}.
As illustrated in Fig.~\ref{fig4}(b), the squeezed-thermal reservoir fundamentally reshapes this synchronization landscape, manifesting a significantly enhanced phase-locking and an extremely localized Arnold tongue. Although the quantitative increase in the synchronization measure $S$ is approximately 50\%, the introduce of squeezed reservoir qualitatively improves the frequency selectivity. To understand this boosting effect, we must bridge the macroscopic synchronization measure with the microscopic Liouvillian eigen-spectrum established in Sec.~\ref{sec3}.

In a reservoir lacking squeezing, the external coherent drive $\epsilon$ must compete against an isotropic and macroscopic decoherence rate (on the order of $\gamma_0/2$). Consequently, a weak drive can only induce a diffuse and fragile phase preference. The system's response is easily smeared out by uniform quantum fluctuations, resulting in a broad but shallow Arnold tongue.
In stark contrast, the squeezed reservoir structurally modifies the dissipation channels. By projecting the external drive onto the eigen-space of the un-driven Liouvillian, we observe a profound dynamical asymmetry. The strong squeezing severely clamps the amplitude components (the $y$ and $z$ axes) with massive relaxation rates ($\sim 2\gamma_0 N$), effectively pinning the system onto the limit cycle trajectory and rendering it highly robust against amplitude perturbations. Simultaneously, the reservoir suppresses the decoherence along the $x$-axis, creating a slow manifold with a vanishingly small phase diffusion rate ($\sim \gamma_0/(8N)$).
This extreme separation of timescales provides a nearly dissipation-free channel for the external signal. Even an exceptionally weak drive $\epsilon$ can easily overcome the residual phase diffusion on this slow manifold, establishing a rigid and long-lived coherent phase locking. The highly localized nature of the Arnold tongue in Fig.~\ref{fig4}(b) is therefore not a limitation, but rather a direct consequence of this underlying slow manifold. The system exhibits an increased frequency selectivity because the strict dynamical conditions required to sustain the limit cycle and resonate with the slow manifold strictly filter out off-resonant perturbations. This synergistic interplay between anisotropic dissipation and coherent driving constitutes the algebraic mechanism of squeezing-boosted quantum phase synchronization.

\section{Synchronization blockade by squeezed reservoir} 
\label{sec5}
The preceding discussion is consistent with prior studies that demonstrate the potential of squeezing to enhance quantum synchronization in different situations \cite{manzano_2013_synchronization,weiss_2017_quantumcoherent,shen_2023_enhancing,sonar_2018_squeezing}. Nevertheless, all of these considerations have only concentrated on the relationship between squeezing strength and quantum synchronization, overlooking the crucial role of the squeezing angle in reshaping the system's dynamics. In this section, we reveal a counter-intuitive phenomenon: tuning the squeezing angle does not enhance synchronization but instead drives the system into a near-perfect quantum synchronization blockade. More fundamentally, we transcend the phenomenological observation of coherence quenching and provide a rigorous physical mechanism for this blockade through the algebraic lens of dynamical orthogonality between the driving Hamiltonian and the Liouvillian dissipation subspace.

\subsection{Synchronization blockade}

We first establish the macroscopic manifestation of the synchronization blockade using the steady-state $S$-function. Referring to the definition in Eq. (\ref{eq15}), it is straightforward to deduce that $S(\phi)=0$ for any given value of $\phi$ indicates the presence of synchronization blockade. According to Eq. (\ref{eq16}), this condition requires that both the $x$ and $y$ components of the Bloch vector simultaneously equal zero.
\begin{eqnarray}
\label{eq17}
\gamma-2\gamma_{0}\text{Re}(M)&=&0,\\
\label{eq18}
\Delta+\gamma_{0}\text{Im}(M)&=&0.
\end{eqnarray}
 {Equation (\ref{eq18}) can always be met by modulating the detuning, whereas Eq. (\ref{eq17}) necessitates the adjustment of the squeezing angle $\Phi$. After simplification, we find that Eq. (\ref{eq17}) can be equivalently expressed as
\begin{equation}
\cos\Phi+\coth2r=0.
\label{eq19}
\end{equation}
This equality can only be reached in the asymptotic limit of infinite squeezing strength ($r \to \infty$) for the case $\cos\Phi = -1$. Therefore, in any practical experiment with large but finite $r$, the blockade condition can only be approximately met, leading to a strong suppression rather than a perfect vanishing of synchronization.}

\begin{figure}
  \includegraphics[width=0.5\textwidth]{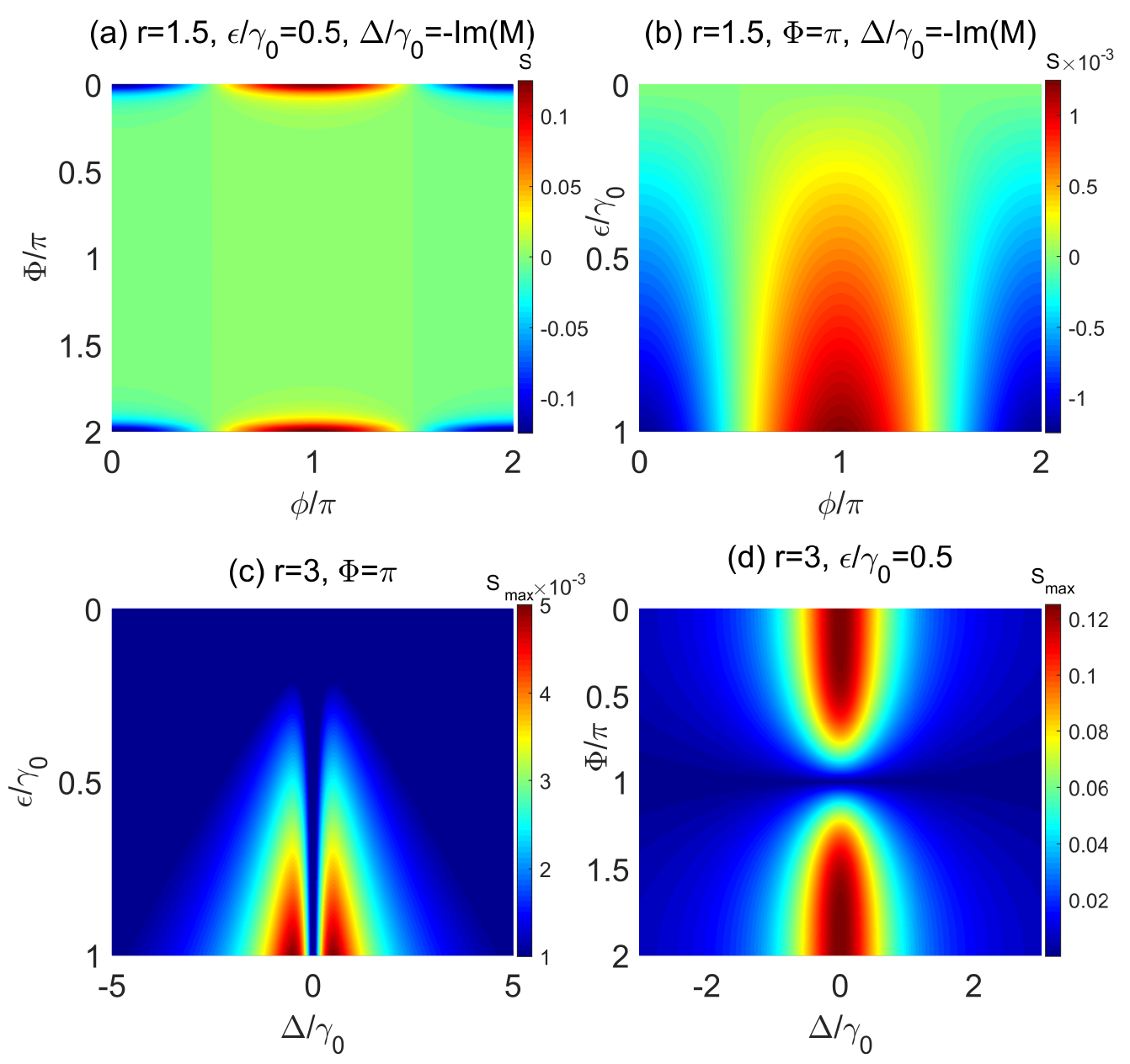}
\caption{(color online) The synchronization blockade in different parameter spaces. (a) $S(\phi)$ as a
function of $\phi$ and $\Phi$ with $r=1.5$, $\epsilon/\gamma_0=0.5$ and $\Delta/\gamma_0=-\text{Im}(M)$.
(b) $S(\phi)$ as a function of $\phi$ and $\epsilon$ with $r=1.5$, $\Phi=\pi$ and $\Delta/\gamma_0=-\text{Im}(M)$.
(c) $S_{\rm max}$ as a function of the signal strength $\epsilon$ and the detuning $\Delta$ with $r=3$ and $\Phi=\pi$.
(d) $S_{\rm max}$ as a function of the squeezing angle $\Phi$ and the detuning $\Delta$ with $r=3$ and $\epsilon/\gamma_0=0.5$.}
\label{fig5}       
\end{figure}

Figure~\ref{fig5}(a) visualizes this mechanism by plotting the synchronization measure $S(\phi)$ in the $\{\Phi, \phi\}$ parameter space, with Eq.~(\ref{eq18}) held true. The squeezing angle $\Phi$ emerges as the critical control parameter governing the dynamical fate of the system: while the system synchronizes strongly at $\Phi=0$ (reproducing the results of Fig.~\ref{fig3}), the phase-locking is systematically suppressed as $\Phi \to \pi$. As dictated by Eq.~(\ref{eq19}), which can only be asymptotically satisfied for large squeezing strength $r$, an ideal blockade ($S(\phi)=0$) is unattainable.

 The consequences of this blockade at $\Phi=\pi$ are further detailed in Figs.~\ref{fig5}(b) and \ref{fig5}(c). 
In sharp contrast to the enhancement of synchronization observed in Figs.~\ref{fig3} and \ref{fig4}, here the synchronization is significantly suppressed. 
Remarkably, the characteristic Arnold tongue structure, though heavily suppressed, persists but is now split at zero detuning ($\Delta=0$), as shown in Fig.~\ref{fig5}(c). 
This splitting is a direct manifestation of the blockade condition in Eq.~(\ref{eq18}), which requires $\Delta=0$ for maximal suppression when $\Phi=\pi$. 
Finally, Fig.~\ref{fig5}(d) presents a global view in the $\{\Phi, \Delta\}$ space, revealing two symmetric Arnold tongues mirrored around $\Phi=\pi$. 
This symmetry is a fundamental consequence of the system's periodic dependence on the squeezing angle $\Phi$ in the master equation, solidifying the role of $\Phi$ as a key control parameter for topologically altering the synchronization landscape.

\subsection{Physical interpretation: dynamical mismatch and coherence quenching}

It is essential to provide an in-depth elucidation of the physical mechanisms underlying the phenomenon of quantum synchronization blockade. While blockade has been observed in other contexts, such as through destructive interference in many-body systems~\cite{wang_2023_absence,kehrer_2024_quantum}, phonon anti-bunching within the same parametric regime \cite{thomas_2022_quantum} or balanced dissipation rates in higher-spin systems~\cite{koppenhfer_2019_optimal,tan_2022_halfinteger}, the mechanism here for a single qubit is fundamentally different. We propose that the synchronization blockade arises because the squeezed reservoir drives the steady-state of the TLS into a classical mixed state, which, by definition, possesses no quantum coherence and thus exhibits no phase preference~\cite{solanki_2023_symmetries,vaidya_2024_exploring}.

To elucidate the microscopic algebraic mechanism underlying this phenomenon, we revisit the analysis of the Liouvillian eigen-spectrum analysis under the un-driven condition. When the squeezing angle is rotated to $\Phi = \pi$, the squeezing parameter undergoes a sign reversal, becoming a positive real number $M = |M|$. Substituting this condition into the diagonalized Liouvillian eigen-equations reveals a dramatic swap in the decay rates of the transverse Bloch components:
\begin{equation}
\lambda_x \approx -2\gamma_0 N,\quad \lambda_y \approx -\frac{\gamma_0}{8N}.
\end{equation}
In the deep quantum regime ($r \gg 1, N \gg 1$), the dissipation structure of the phase space rotates globally by 90 degrees. The exceptionally long-lived slow manifold previously located on the $x$-axis is transferred to the $y$-axis, while the $x$-axis instantly degrades into a fast dissipation channel subjected to severe quantum fluctuations.

This rotation of the phase-space dissipation landscape directly dictates the failure of the external weak drive. In our physical model, the semiclassical driving Hamiltonian is generated by the operator $\hat{\sigma}_y$, as depicted in Eq. (\ref{eq4}). Governed by the Lie algebra commutation relation $[\hat{\sigma}_y, \hat{\sigma}_z] = 2i\hat{\sigma}_x$, this specific weak external field exclusively injects quantum coherence along the $x$-axis.
This underlying algebraic structure reveals the deepest physical mechanism of the quantum synchronization blockade: a  dynamical mismatch. When $\Phi=\pi$, the direction in which the external drive attempts to establish coherence (the $x$-axis) is precisely the fastest dissipation channel. Any weak coherence injected by the external field is instantaneously eradicated by the massive decay rate of $2\gamma_0 N$, undergoing rapid coherence quenching. Simultaneously, although the $y$-axis now serves as the long-lived slow manifold, the driving field $\hat{\sigma}_y$ commutes with itself ($[\hat{\sigma}_y, \hat{\sigma}_y] = 0$), exerting absolutely zero coherence on this dissipation-free dimension.
This dynamical orthogonality between the external driving Hamiltonian and the Liouvillian slow manifold fundamentally locks the system, preventing the accumulation of any steady-state off-diagonal elements ($r_x \to 0, r_y \to 0$).

 The physical mechanism becomes transparent when considering the mathematical foundation of the synchronization measure itself. The $S$-function in Eq.~(\ref{eq16}) is a linear combination of the Bloch vector components $r_x$ and $r_y$, which fully determine the off-diagonal elements of the qubit's density matrix. 
To gain an intuitive understanding of the interplay between quantum coherence and synchronization blockade, we utilize the $l_{1}$ norm as a measure of coherence \cite{baumgratz_2014_quantifying}: 
\begin{equation}
C_{l_{1}}=\sum_{i\neq j}|\rho_{ij}|=2|\rho_{01}|=\sqrt{r^{2}_{x}+r^{2}_{y}}.
\end{equation}
Consequently, the condition for an ideal synchronization blockade, $r_x = r_y = 0$, is mathematically identical to the condition for vanishing quantum coherence, $C_{l_1} = 0$. Synchronization is therefore intrinsically linked to coherence; the former cannot exist without the latter.

\begin{figure}
  \includegraphics[width=0.48\textwidth]{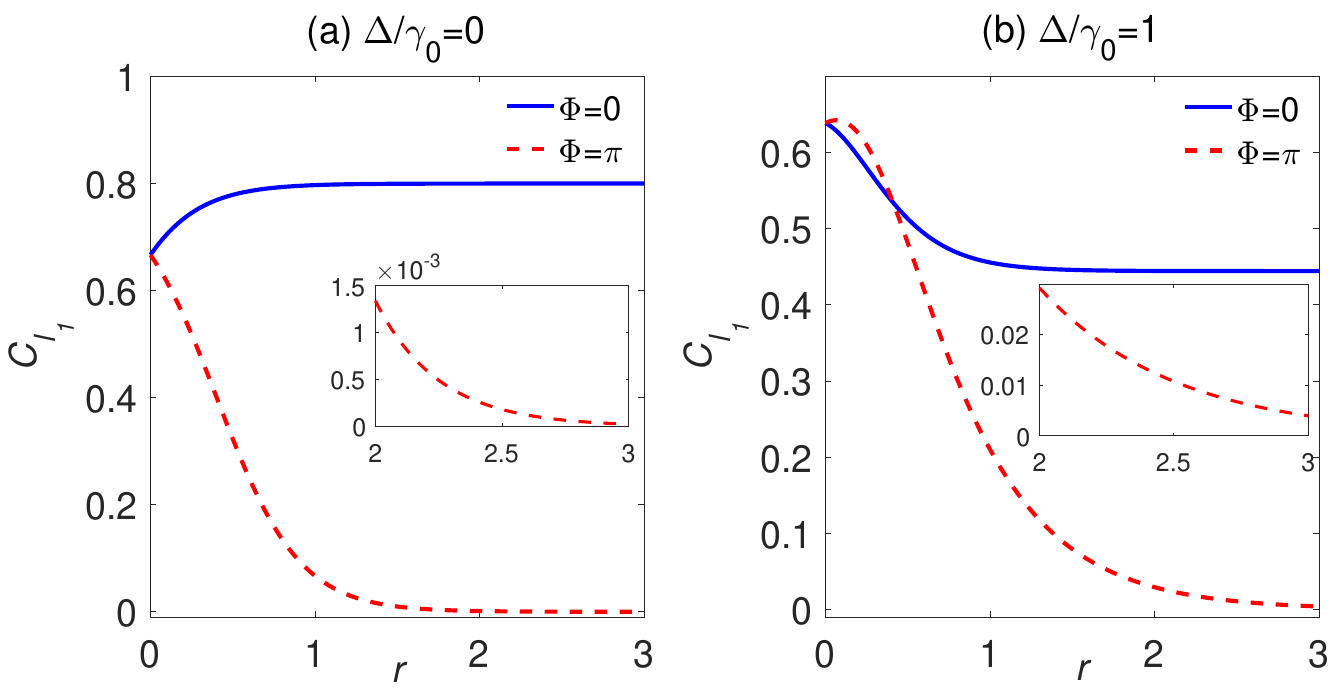}
\caption{(color online) The steady-state coherence as a function of the squeezing strength for different values of the squeezing angle. (a) $\Delta/\gamma_0=0$ and (b) $\Delta/\gamma_0=1$. The other parameters are $\epsilon/\gamma_0=1$ and $n=0$. The insets provide a detailed view of the steady-state coherence within the region of $r\in[2,3]$.}
\label{fig6}       
\end{figure}

The steady-state coherence evolution based on the $l_1$ norm in Fig.~\ref{fig6} provides direct numerical verification for this algebraic deduction. As shown in Fig.~\ref{fig6}(a) for the resonant case ($\Delta=0$), the behavior of coherence critically depends on the squeezing angle $\Phi$. When $\Phi=0$, corresponding to the boosting regime, coherence is enhanced with increasing $r$, which enables the stronger synchronization observed in Sec.~\ref{sec4}. Conversely, for $\Phi=\pi$, the system is driven towards the blockade regime; coherence is rapidly quenched as $r$ increases, forcing the TLS into a classical mixed state and thereby suppressing any phase-locking with the external drive.

Furthermore, Fig.~\ref{fig6}(b) reveals the nuanced role of detuning. In the boosting regime ($\Phi=0$), a non-zero detuning is detrimental, reducing the steady-state coherence and thus weakening synchronization, consistent with the behavior of the Arnold tongue in Fig.~\ref{fig4}(b), where $S_{\rm max}$ decreases with increasing detuning. In the blockade regime ($\Phi=\pi$), however, a modest detuning can partially mitigate the coherence decay induced by the reservoir, as shown in the insets of Fig.~\ref{fig6}. This effect explains the split Arnold tongue structure observed in Fig.~\ref{fig5}(c), where synchronization is marginally recovered for finite $\Delta$, thereby hindering the onset of quantum synchronization blockade.

\section{Experimental Discussion} 
\label{sec6}
Quantum synchronization has been observed in a number of different systems, including IBM's quantum cloud computing system \cite{koppenhfer_2020_quantum}, the $^{87}$Rb cold atom system \cite{laskar_2020_observation}, nuclear spin pairs in nuclear magnetic resonance (NMR) system \cite{krithika_2022_observation}, the circuit QED system \cite{nigg_2018_observing,tao_2024_noiseinduced}, and the trapped-ion system \cite{zhang_2023_quantum}. The model considered in this paper is relatively straightforward and can be implemented in a number of different physical systems.
Here, we propose an experimental proposal to demonstrate the enhancement of quantum phase synchronization through the application of squeezed-reservoir engineering in a circuit QED system. We consider a superconducting transmon qubit that is coupled to a common electromagnetic field in a waveguide. At present, the squeezed-vacuum reservoir can be generated by injecting the waveguide with a broadband squeezed field. The transmon qubit is driven by a semiclassical signal of strength $\epsilon$ at frequency $\omega_{L}$. Therefore, in principle, the driven-dissipative model described by Eq. (\ref{eq5}) can be obtained. 

The fundamental technology underpinning the proposed scheme resides in the engineering of the squeezed reservoir. Whether one aims to observe the squeezing enhanced quantum phase synchronization or to explore the squeezing induced quantum synchronization blockade, a highly squeezed reservoir is imperative. Consequently, the first experimental consideration is that of the squeezing strength $r$. According to the conversion relation $r_{\rm dB}=r\times20/\log(10)$, $r=1.5$ corresponds to a 13 dB squeezed light. In Refs. \cite{vahlbruch_2016_detection,macklin_2015_a}, the generation of 15 dB squeezed light and 20 dB squeezed microwave has been demonstrated through the use of optical parametric down conversion and Josephson parametric amplification, respectively. The second consideration is how to modulate the squeezing angle $\Phi$. In the experiment, the squeezing angle is determined by the relative phase between the squeezed light and a strong coherent reference beam called the local oscillator (LO) \cite{schnabel_2017_squeezed}. One can use electro-optic modulators (EOMs) to dynamically control the phase of the LO, allowing real-time adjustment of the squeezing angle \cite{chelkowski_2005_experimental}. We believe that the aforementioned techniques are perfectly suited to the experimental requirements of this study.

From an experimental perspective, the theoretical phase-space distributions (e.g., the Husimi $Q$-function) and synchronization measures (e.g., the $S$-function) evaluated in our model are firmly grounded in accessible physical observables. The definitive Hermitian observables required to distinguish genuine phase synchronization from a passive forced response are the transverse Pauli operators, $\hat{\sigma}_x$ and $\hat{\sigma}_y$. Because the TLS is continuously driven, an unsqueezed environment still yields a finite forced response ($\langle\hat{\sigma}_{x,y}\rangle \neq 0$). However, this passive forced response decays trivially (e.g., following a Lorentzian profile) as the frequency detuning $\Delta$ increases, and its relative phase drifts linearly, as shown in Fig.~\ref{fig4}(a). In stark contrast, true synchronization is physically distinguished by a macroscopic enhancement in the transverse polarization amplitude $P_\perp = \sqrt{\langle\hat{\sigma}_x\rangle^2 + \langle\hat{\sigma}_y\rangle^2}$. More crucially, genuine phase entrainment locks both this transverse amplitude and the relative phase angle $\phi = \arctan(\langle\hat{\sigma}_y\rangle/\langle\hat{\sigma}_x\rangle)$ within a specific detuning window, as shown in Fig.~\ref{fig4}(b). By scanning the detuning and measuring these observables, one can experimentally map out the Arnold tongue—the ultimate hallmark of limit-cycle entrainment. In contemporary quantum platforms, such as trapped ions or superconducting circuits, these Hermitian observables are routinely extracted via quantum state tomography. Indeed, recent experiments have successfully utilized exactly these projective Pauli measurements to reconstruct the density matrix and unequivocally demonstrate single-qubit quantum phase synchronization \cite{zhang_2023_quantum}.

\section{Conclusion}
\label{sec7}
In summary, we have performed a theoretical investigation into the quantum phase synchronization of a TLS interacting with a squeezed reservoir, revealing its dual role. By deriving the TLS model from an oscillator and analyzing its extreme dissipation anisotropy via the Liouvillian eigen-spectrum,
we first established that the squeezed reservoir induces a stable limit cycle, transforming the passive TLS into a self-sustained quantum oscillator. This enables a qualitative transition from the weak forced response characteristic of dissipative systems to robust, high-quality synchronization. This enhancement manifests in two key ways: a greater synchronization strength (higher $S_{\rm max}$) and an increased frequency selectivity.

More strikingly, we demonstrated that the squeezing angle acts as a powerful control knob to actively suppress synchronization. By tuning this angle towards $\pi$, the reservoir drives the system into a classical mixed state with vanishing quantum coherence, thus inducing a profound synchronization blockade. The physical origin of the blockade was directly traced to the quenching of the off-diagonal elements of the density matrix. We have proposed a feasible experimental implementation within the circuit QED framework, underscoring the practical potential of our findings. This work establishes squeezed-reservoir engineering not merely as a method for noise suppression, but as a sophisticated strategy for dynamically modulating the quantum phase synchronization.

Future investigations may delve into the transient Liouvillian dynamics associated with switching between the boosting and blockade regimes. Extending this active control scheme to small quantum networks could enable the topological routing of phase information, where individual nodes are dynamically activated or silenced via local reservoir engineering. Furthermore, while controlling the synchronization of a TLS using external signals through squeezed-reservoir engineering constitutes an initial step, a more significant area for in-depth exploration lies in the effects of squeezed-reservoirs on the spontaneous and mutual synchronization of two qubits, similar to macroscopic phase locking observed in many-body mixed states \cite{buca_2022_algebraic,tindall_2020_quantum}. We anticipate that two interacting qubits, each immersed in an independent squeezed reservoir, can spontaneously break the continuous phase symmetry and achieve robust macroscopic phase locking even in the strict absence of external coherent driving. Physically, this collective phenomenon would be driven by the non-local coherences pumped synergistically by the inter-qubit coupling and the anomalous squeezed vacuum fluctuations. However, the ways in which different types of interactions may block synchronization remain an exciting avenue for future research.

\begin{acknowledgements}
Y. L. Li is supported by National Natural Science Foundation of China under Grant No. 12365003 and Jiangxi Provincial Key Laboratory of Multidimensional Intelligent Perception and Control of China (No. 2024SSY03161). X. Xiao is supported by National Natural Science Foundation of China under Grant No. 12265004 and
Jiangxi Provincial Natural Science Foundation under Grant No. 20242BAB26010. T. X. Lu is supported by the National Natural Science Foundation of China under Grant Nos. 12205054, 12565001 and the Natural Science Foundation of Jiangxi Province under Grant No. 20252BAC200163.
\end{acknowledgements}

\section*{DATA AVAILABILITY}  
The data that support the findings of this article are not publicly available. The data are available from the authors upon reasonable request.

\appendix

\section{Derivation Eq. (\ref{eq5}) from the Bosonic Quantum Stuart-Landau Oscillator}
\label{sec:QSLO}
We demonstrate that the driven TLS subjected to a squeezed reservoir, described by Eq. (\ref{eq5}) in the main text also emerges as the exact deep quantum limit of a bosonic continuous variable oscillator exhibiting strong Kerr non-linearity, immersed in a squeezed-thermal bath. This derivation explicitly bridges our model with the paradigm of the QSLO.

Considering a single-mode bosonic oscillator driven by a classical coherent field and coupled to a squeezed-thermal reservoir. To introduce a mechanism for amplitude saturation, we incorporate a Kerr nonlinearity. In the frame rotating at the external drive frequency $\omega_L$, the Hamiltonian of this bosonic system is given by
\begin{equation}
\hat{H}_{B} = \Delta \hat{a}^\dagger \hat{a} + K (\hat{a}^\dagger \hat{a})^2 + i\frac{\epsilon}{2}(\hat{a} - \hat{a}^\dagger), \label{A1}
\end{equation}
where $\hat{a}$ ($\hat{a}^\dagger$) is the bosonic annihilation (creation) operator, $\Delta = \omega_0 - \omega_L$ is the detuning between the cavity resonance and the drive, $K$ dictates the strength of the Kerr nonlinearity, and $\epsilon$ represents the drive amplitude.The complete open system dynamics are governed by the master equation:
\begin{equation}
\frac{d\rho}{dt} = -i[\hat{H}_{B}, \rho] + \mathcal{L}_{s}\rho,
\end{equation}
where the squeezed-thermal dissipator $\mathcal{L}_{s}$ is defined in terms of standard Lindblad superoperators $\mathcal{D}[\hat{O}]\rho = 2\hat{O}\rho\hat{O}^\dagger - \hat{O}^\dagger\hat{O}\rho - \rho\hat{O}^\dagger\hat{O}$ as:\begin{eqnarray}
\mathcal{L}_{s}\rho &=& \frac{1}{2}\gamma_0(N+1)\mathcal{D}[\hat{a}]\rho + \frac{1}{2}\gamma_0 N \mathcal{D}[\hat{a}^\dagger]\rho \nonumber\\
&& - \gamma_0 M \hat{a}^\dagger \rho \hat{a}^\dagger - \gamma_0 M^* \hat{a} \rho \hat{a}. 
\label{A3}
\end{eqnarray}
Here, $\gamma_0$ is the dissipation rate, and $N$ and $M$ are the standard squeezed reservoir parameters.

The critical transition to the TLS model relies on analyzing the energy spectrum of the un-driven oscillator, $E_n = n\Delta + K n^2$. The transition frequency between the ground and first excited states is $\omega_{0 \to 1} = \Delta + K$, whereas the subsequent transition shifts significantly to $\omega_{1 \to 2} = \Delta + 3K$. In the strong Kerr limit ($K \to \infty$), the non-linearity $2K$ becomes the dominant energy scale in the system. Provided that the drive amplitude $\epsilon$ and the dissipation rate $\gamma_0$ are sufficiently weak compared to $2K$, transitions to higher-order Fock states are strictly forbidden due to the pronounced energy blockade. Consequently, the infinite-dimensional bosonic Hilbert space is naturally truncated to a two-dimensional subspace spanned by $\{|0\rangle, |1\rangle\}$.

Within this restricted two-level approximation, the bosonic operators elegantly map onto the SU(2) Pauli algebra:
\begin{equation}\hat{a} \to \hat{\sigma}_{-}, \quad \hat{a}^\dagger \to \hat{\sigma}_{+}, \quad \hat{a}^\dagger \hat{a} \to \frac{1}{2}(\hat{\sigma}_{z} + \mathbb{I}).
\end{equation}
Applying this exact algebraic projection to the master equation (\ref{A3}) yields profound simplifications. The number operator maps as $\Delta \hat{a}^\dagger \hat{a} \to \frac{\Delta}{2}\hat{\sigma}_z$ by ignoring the identity matrix $\frac{\Delta}{2}\mathbb{I}$. The coherent drive translates seamlessly into a spin rotation: $i\frac{\epsilon}{2}(\hat{a} - \hat{a}^\dagger) \to i\frac{\epsilon}{2}(\hat{\sigma}_{-} - \hat{\sigma}_{+}) = \frac{\epsilon}{2}\hat{\sigma}_y$. Remarkably, substituting the Pauli operators into the squeezed-thermal dissipator directly expands to the precise dissipative terms of our squeezed TLS model. For instance, the standard photon loss and thermal pumping terms map respectively to $-\frac{1}{2}\gamma_0(N+1)(\hat{\sigma}_{+}\hat{\sigma}_{-}\rho + \rho\hat{\sigma}_{+}\hat{\sigma}_{-} - 2\hat{\sigma}_{-}\rho\hat{\sigma}_{+})$ and $-\frac{1}{2}\gamma_0 N(\hat{\sigma}_{-}\hat{\sigma}_{+}\rho + \rho\hat{\sigma}_{-}\hat{\sigma}_{+} - 2\hat{\sigma}_{+}\rho\hat{\sigma}_{-})$. The phase-sensitive anomalous squeezed correlations also survive this projection: $-\gamma_0 M \hat{a}^\dagger \rho_B \hat{a}^\dagger \to -\gamma_0 M \hat{\sigma}_{+} \rho \hat{\sigma}_{+}$. Combining these mapped terms, we rigorously recover the exact TLS master equation utilized in our study:
\begin{eqnarray}
\label{eq:appendix_master}
\frac{d\rho}{dt}&=&-\frac{i}{2}[\Delta\hat{\sigma}_{z}+\epsilon\hat{\sigma}_{y},\rho]-\gamma_{0}M\hat{\sigma}_{+}\rho\hat{\sigma}_{+}-\gamma_{0}M^{*}\hat{\sigma}_{-}\rho\hat{\sigma}_{-}\nonumber\\
&&-\frac{1}{2}\gamma_{0}(N+1)(\hat{\sigma}_{+}\hat{\sigma}_{-}\rho+\rho\hat{\sigma}_{+}\hat{\sigma}_{-}-2\hat{\sigma}_{-}\rho\hat{\sigma}_{+})\nonumber\\
&&-\frac{1}{2}\gamma_{0}N(\hat{\sigma}_{-}\hat{\sigma}_{+}\rho+\rho\hat{\sigma}_{-}\hat{\sigma}_{+}-2\hat{\sigma}_{+}\rho\hat{\sigma}_{-}).
\end{eqnarray}
This derivation unambiguously clarifies the phenomenological mapping between a standard QSLO and our TLS model in the deep quantum limit.

\section{Stability analysis of the un-driven system}
\label{sec:appendix_stability}

In this appendix, we provide a rigorous analytical study of the stability of un-driven ($\epsilon = 0$) TLS. The goal is to formally demonstrate why a vacuum reservoir leads to a single stable ground state, while a squeezed reservoir gives rise to a stable limit cycle, thus providing the theoretical foundation for the numerical results in Fig.~\ref{fig1}.

\subsection*{B.1 Equations of Motion}

In the absence of an external drive ($\epsilon=0$), the dynamics of the TLS can be described using spherical coordinates for the Bloch vector, $\theta$ and $\phi$, as given by Eqs.~(\ref{eq10}) and (\ref{eq11}) in the main text:
\begin{align}
    \frac{d\theta}{dt} &= f(\theta) = \gamma_0 \csc\theta + \gamma \cot\theta \label{eq:a1}, \\
    \frac{d\phi}{dt} &= \omega_0 + \gamma_0 \text{Im}(M)\cos(2\phi) + \gamma_0 \text{Re}(M)\sin(2\phi). \label{eq:a2}
\end{align}
Here, $\gamma = \gamma_0(2N+1)$. We will now analyze the stability for two distinct cases.

\subsection*{B.2 Case I: Vacuum Reservoir ($r=0, n=0$)}

For a vacuum reservoir at zero temperature, the squeezing parameter $r=0$ and the thermal photon number $n=0$. This leads to $N=0$ and $M=0$. Consequently, the dissipation rate $\gamma = \gamma_0$.

The equation for $\theta$ dynamics (\ref{eq:a1}) simplifies to:
\begin{equation}
    \frac{d\theta}{dt} = \gamma_0 (\csc\theta + \cot\theta) = \gamma_0 \cot(\theta/2).
\end{equation}
A steady state, or fixed point ($\dot{\theta}=0$), is found where $\cot(\theta/2)=0$. This equation has a unique solution in the physically relevant range $\theta \in [0, \pi]$, which is $\theta_s = \pi$. This corresponds to the ground state of the TLS, i.e., the south pole of the Bloch sphere $\vec{r} = (0, 0, -1)$.

To analyze the stability of this fixed point, we consider a small perturbation $\delta\theta$ away from it, such that $\theta = \pi - \delta\theta$, with $\delta\theta > 0$. The rate of change of the perturbation is given by:
\begin{equation}
    \frac{d(\pi - \delta\theta)}{dt} = \gamma_0 \cot\left(\frac{\pi - \delta\theta}{2}\right) = \gamma_0 \tan\left(\frac{\delta\theta}{2}\right).
\end{equation}
For a small perturbation, $\tan(\delta\theta/2) \approx \delta\theta/2$. The linearized equation becomes:
\begin{equation}
    \frac{d(\delta\theta)}{dt} \approx -\frac{\gamma_0}{2} \delta\theta.
\end{equation}
Since the coefficient $-\gamma_0/2$ is negative, any small perturbation $\delta\theta$ will decay exponentially to zero. This proves that the fixed point at $\theta_s = \pi$ is stable. Therefore, in a vacuum reservoir, the TLS will always decay to its ground state, and no self-sustained oscillation (limit cycle) can exist.

\subsection*{B.3 Case II: Squeezed Reservoir ($r>0$)}

When the reservoir is squeezed ($r>0$), the parameters $N$ and $M$ are non-zero. We first find the steady state for the $\theta$ dynamics by setting Eq.~(\ref{eq:a1}) to zero:
\begin{equation}
    \gamma_0 \csc\theta + \gamma \cot\theta = \frac{\gamma_0 + \gamma \cos\theta}{\sin\theta} = 0.
\end{equation}
This yields a steady-state angle $\theta_s$ that satisfies:
\begin{equation}
    \cos\theta_s = -\frac{\gamma_0}{\gamma} = -\frac{1}{2N+1}.
\end{equation}
This is the same result as Eq.~(\ref{eq12}). For any $r>0$, we have $N > 0$, which means $\theta_s$ is a valid angle between $\pi/2$ and $\pi$. Crucially, this condition does not constrain $\phi$. This implies that there is not an isolated fixed point, but rather a continuum of steady states forming a circle on the Bloch sphere at a constant polar angle $\theta = \theta_s$.

Next, we analyze the stability of this manifold of states. We linearize the function $f(\theta)$ from Eq.~(\ref{eq:a1}) around $\theta_s$. The stability is determined by the sign of the derivative $\frac{df}{d\theta}$ at $\theta_s$:
\begin{equation}
    \frac{df}{d\theta}\bigg|_{\theta_s} = \frac{-\gamma\sin^2\theta_s - (\gamma_0 + \gamma\cos\theta_s)\cos\theta_s}{\sin^2\theta_s}.
\end{equation}
Since $\gamma_0 + \gamma\cos\theta_s = 0$ at the steady state, the expression simplifies to:
\begin{equation}
    \frac{df}{d\theta}\bigg|_{\theta_s} = \frac{-\gamma\sin^2\theta_s}{\sin^2\theta_s} = -\gamma_0(2N+1).
\end{equation}
Because this derivative is negative, the manifold is stable in the $\theta$ direction. Any trajectory starting off this circle will be attracted towards it.

Finally, the dynamics along this circle are governed by the $\phi$ equation (\ref{eq:a2}). As long as $M \neq 0$, $\dot{\phi}$ is generally non-zero and depends on $\phi$, causing the system's state to perpetually evolve along the stable circle $\theta=\theta_s$.

Such a stable manifold that attracts nearby trajectories and possesses internal dynamics, is precisely a stable limit cycle. Our analysis rigorously confirms that a squeezed reservoir is essential for creating the self-sustained oscillation necessary for synchronization.



\end{document}